\documentclass[preprint]{revtex4}
\usepackage{amsmath,amssymb,amsbsy,latexsym}
\usepackage{hyperref}
\usepackage{color}

\def\CQ{{\mathcal Q}}
\def\CR{{\mathcal R}}

\def\BC{{\mathbb C}}
\def\BR{{\mathbb R}}
\def\BZ{{\mathbb Z}}
\def\FS{{\mathfrak S}}
\def\tr{{\rm Tr}}

\begin{document}

\title{A Classical Solution in Six-dimensional Gauge Theory\\ with Higher Derivative Coupling}

\author{Hironobu Kihara\footnote{ 
Current address: School of Physics, Korea Institute for Advanced Study,
207-43 Cheongnyangni-2dong, Dongdaemun-gu, Seoul 130-722, Korea
}}
\email{hkihara(at)kias.re.kr}
\affiliation{ Osaka City University, Advanced Mathematical Institute (OCAMI),
3-3-138 Sugimoto, Sumiyoshi, Osaka 558-8585, Japan}
\author{Muneto Nitta}
\email{nitta(at)phys-h.keio.ac.jp}
\affiliation{Department of Physics, Keio University, Hiyoshi, Yokohama, Kanagawa
 223-8521, Japan
}

\preprint{OCU-PHYS 261}
\preprint{hep-th/yymmnnn}

\begin{abstract}
We show that the spin connection of the standard metric on 
a six-dimensional sphere gives an exact solution 
to the generalized {self-duality} equations 
suggested by Tchrakian some years ago. 
We work on an SO(6) gauge theory with a higher-derivative coupling term. 
The model consists of vector fields only. The pseudo-energy is 
bounded from below
by a topological charge 
which is proportional to the winding number of spatial $S^5$ around the internal space SO(6). The fifth homotopy group of SO(6) is, indeed, $\BZ$. The coupling constant of higher derivative term is quadratic in the radius of the underlying space $S^6$. 
\end{abstract}

\maketitle

Classical solutions to field equations often play 
an important role in study of non-perturbative effects 
in field theory or string theory.
Instantons are such objects. 
Yang-Mills instantons on $\BR^4$ or on a four sphere $S^4$
found by Belavin et al. (BPST) \cite{Belavin:1975fg} 
are topological solitons 
classified by a map from the spatial boundary 
(three-dimensional sphere $S^3$) to a gauge group SU(2);
the third homotopy group is 
$\Pi_3 [{\rm SU(2)}] \simeq \BZ$. 
Since its discovery the Yang-Mills instantons have been 
studied by many people in both physics and mathematics. 
One important achievement is the determination 
of low energy effective action of ${\cal N}=2$ supersymmetric gauge theories 
\cite{Seiberg:1994rs}.
String theory or M-theory is considered 
to unify space-time, matter and all forces including gravity, 
but it requires ten or eleven dimensional space-time 
contrary to our four dimensional space-time. 
Two different mechanisms 
to compactify extra space in higher dimensions
have been suggested and studied extensively; 
the Kaluza-Klein compactification \cite{Nordstrom:1988fi,Kaluza:1921tu} and 
brane world scenarios \cite{brane-world}.
Classical solutions (solitons) to field equations 
play important roles in both scenarios: 
in the former non-trivial field configurations on a compact space 
are used for spontaneous compactification \cite{Cremmer:1976ir},   
and in the latter solitons themselves as extended objects 
may realize four-dimensional world 
on their world-volume.

Some years ago Tchrakian suggested a broad class of 
higher-dimensional generalization of 
instantons and monopoles \cite{Tchrakian:1978sf}. 
Since then several people have been studying these 
generalized self-duality equations. 
Tchrakian constructed the BPST analogues in all $4p$ dimension on $\BR^{4p}$ \cite{Tchrakian:1984gq}.  
Due to the scale invariance of these systems, 
it is also possible to map the systems on $\BR^{4p}$ onto the spheres 
$S^{4p}$ \cite{O'Se:1987fx}. 
The solitons on the complex projective space $\BC{\rm P}^n$ 
were also considered \cite{Ma:1990ja}.
The Grossman-Kephart-Stasheff solution on 
an eight-dimensional sphere is an example of 
an exact solution to the generalized self-duality equations 
\cite{Grossman:1984pi}. 
In \cite{Chen:1999ab} non-Abelian Berry's phases suggest the localized monopoles with codimensions greater than five in ten-dimensional space-time. 
Finite energy solution in six-dimensional Minkowski space has been recently studied numerically \cite{Kihara:2004yz,Radu:2005rf}. 

In this article we construct a new exact instanton 
solution to the generalized 
self-duality equations with gauge group SO(6) 
on a six-dimensional sphere $S^6$. 
Our solution saturates the Bogomol'nyi type bound 
for the pseudo-energy and so is stable.
Since the sphere $S^6$ is covered by two patches each of which 
is diffeomorphic to $\BR^6$,
the topological charge is given by a map from 
a spatial boundary of one patch (five-dimensional sphere $S^5$) 
to the gauge group SO(6),
as the generalization of BPST instantons on $S^4$. 
This map is classified by the fifth homotopy group, 
$\Pi_5 [{\rm SO(6)}] \simeq \BZ$. 
A six-dimensional compact space is particularly important because 
a soliton on it may give spontaneous compactification 
from ten dimensions, suggested by string theory, to four dimensions.
In fact we see that the radius of $S^6$ is related to 
the pseudo energy of our solution, 
suggesting the stabilization of the modulus 
of the compactified space.
We consider an SO(6) gauge theory with 
a higher derivative coupling. 
Several people study higher derivative coupling terms in the context of D-branes as higher derivative corrections to a non-Abelian Dirac-Born-Infeld soliton \cite{Bergshoeff:1986jm}. 
We expect that our higher derivative coupling has such a origin.

Our model consists of gauge fields $A_{i}^{[ab]}$ $(i,a,b=1,\cdots,6)$. 
The group SO(6) is locally isomorphic to SU(4). 
{We use the Clifford algebra of SO(6) for 
the representation of gauge group.}\footnote{
 {The reason why we use the Clifford algebra is 
that the pseudo-energy defined below in 
Eq.~(\ref{pseudo-energy}) 
can be expressed in a single trace term. 
The pseudo-energy expressed in SO(6) indices 
equivalent to ours cannot be expressed in a single trace 
and is very complicated.
}
}

 {It} 
is generated by  {the gamma matrices} 
$\Gamma_{a}$, where $a=1, \cdots, 6$. These generators $\{ \Gamma_a \}$ are realized as Hermitian $8 \times 8$ matrices with complex coefficients. Thus the Lie algebra is embedded in the su(8) algebra.   They satisfy the anti-commutation relation $\{ \Gamma_a , \Gamma_b  \}=2\delta_{ab}$ and their 
 {commutators} 
$\displaystyle \Gamma_{ab} = \frac{1}{2}[ \Gamma_a , \Gamma_b ]$ generate the Lie algebra so(6).  The chirality operator $\gamma_7 = - i \Gamma_1 \cdots \Gamma_6$ plays an important role. The operator $\gamma_7$ anti-commutes with $\Gamma_{a}$ and commutes with $\Gamma_{ab}$.
 { 
$\Gamma_a$'s and $\gamma_7$ are Hermitian 
($\Gamma_a^{\dag} =\Gamma_a$, $\gamma_7^{\dag} =\gamma_7$) 
whereas $\Gamma_{ab}$'s are anti-Hermitian.
} 
 Antisymmetric products of $\Gamma$'s are defined as 
\begin{align}
 \Gamma_{a_1 \cdots a_s} =\frac{1}{s!} \sum_{\sigma \in \FS_s} {\rm sign}(\sigma) \Gamma_{a_{\sigma(1)}} \cdots \Gamma_{a_{\sigma(s)}}~~, 
\end{align}
 where the sum is taken over all permutations of $\{1,\cdots,  s\}$. 
 {A set of} 
the matrices $\{1,\Gamma_a, \Gamma_{ab},\Gamma_{abc},
\Gamma_{ab}\gamma_7,\Gamma_a \gamma_7, \gamma_7\}$ 
generates the total Clifford algebra. Thus the dimension of the Clifford algebra is equal to $2^6=64$. 
The Clifford algebra can be graded by the number of $\Gamma_a$'s where $a$ runs from 1 to 6. For instance the grade of the generator $\Gamma_{ab}\gamma_7$ is fourth.   
The (anti-)commutation relations of $\Gamma_{ab}$'s are given as follows,
\begin{align}
\{ \Gamma_{ab} ,  \Gamma_{cd} \}  &= 2 \left( \delta_{bc}\delta_{ad} - \delta_{bd}\delta_{ac} + \Gamma_{abcd} \right)  ~,\\
[ \Gamma_{ab} , \Gamma_{cd} ] &= 2 \left( \delta_{bc} \Gamma_{ad} - \delta_{bd} \Gamma_{ac} - \delta_{ac} \Gamma_{bd} + \delta_{ad} \Gamma_{bc}   \right)~~.
\end{align}
The anti-commutation relations imply that $\tr \Gamma_{ab} \Gamma_{cd}= -8 (\delta_{ac}\delta_{bd} - \delta_{ad}\delta_{bc}) = -8 \delta^{ab}_{[cd]}$. The matrices $\Gamma_{abcd}$ is related to $\Gamma_{ef}$ by the chiral operator $\gamma_7$ and totally antisymmetric tensor $\epsilon_{abcdef}$: 
\begin{align}
\Gamma_{abcd} &= - \frac{i}{2!}  \gamma_7 \epsilon_{abcdef}  \Gamma_{ef}~,
\end{align}
where $\epsilon$ is normalized as $\epsilon_{123456}=1$. 

The base space  {which we consider} is a six-dimensional sphere $S^6$. 
  {The sphere $S^6$ is covered by 
the two patches which are diffeomorphic to $\BR^6$. 
We work on one of them.}
 We denote the space  {coordinates by} $x^i$. 
 {We consider the standard metric on the} sphere: 
the metric and curvature tensors are  
\begin{align}
ds^2 &= \frac{\delta_{ij}}{(1+x^2/4R_0^2)^2} dx^{i} dx^{j}= g_{ij}dx^{i} dx^{j} ~~, & R^{i}_{jkl} &=\frac{1}{R_0^2} \left( \delta^i_k g_{jl} - \delta^i_l g_{jk}   \right)~,\cr
 \CR_{ij} &= \frac{5}{R_0^2}g_{ij}~,& \CR &= \frac{30}{R_0^2}~~,
\end{align}
where the parameter $R_0$ is the radius. 
The determinant of the metric is $g=\det{g_{ij}}= (1+x^2/4R_0^2)^{-12}$. 
We often abbreviate the basis of forms; $dx^{i_1} \wedge \cdots \wedge dx^{i_k}=dx^{i_1 \cdots i_k}$. 
The six-form $dv=dx^{1\cdots 6} \sqrt{g}$ is an invariant volume form. Indeed the form is invariant under general coordinate transformations.  The integration $\int_{S^6} dv$ gives the volume of $S^6$. 
 The Hodge dual of a base of differential forms, $dx^{i_1 \cdots i_s}$, is given as 
\begin{align}
 * dx^{i_1 \cdots i_s} = \frac{1}{(6-s)!}\frac{1}{\sqrt{g}}  \epsilon^{i_1 \cdots i_s j_1 \cdots j_{6-s}} g_{j_1j'_1 } \cdots g_{j_{6-s}j'_{6-s}} dx^{j'_1 \cdots j'_{6-s}}~~,
\end{align}
 where $\epsilon^{ijklmn}$ is again a totally antisymmetric tensor and is normalized as $\epsilon^{123456}=1$. 
One must take care of the position of the indices which  {should} be
raised and lowered by the metric tensor.
  The  use of  {sechsbein} $e^{I}=dx^I/(1+x^2/4R_0^2),~(I=1,2,\cdots,6)$ makes notation simple.  
They form an orthonormal frame, $ds^2 = \delta_{IJ} e^I e^J$. Thus we can relax the index position of $I,J$. The spin connection is defined by the relation; $de^I= - \omega^{IJ} e^J$. One can easily see that $\omega^{IJ} = (x^I e^J - x^J e^I)/2R_0^2$. 
 The Riemann curvature is expressed as $R^{IJ}= d \omega^{IJ} + \omega^{IK} \wedge \omega^{KJ}= e^I\wedge e^J / R_0^2$. We often mix up these indices with the internal indices $a,b,\cdots$.  The Hodge dual of their products becomes
\begin{align}
 * e^{I_1} \wedge \cdots \wedge e^{I_s} &= \frac{1}{(6-s)!} \epsilon^{I_1 \cdots I_6} e^{I_{s+1}}  \wedge \cdots \wedge e^{I_6}~~.
\end{align}

The dynamical degree of freedom of our model is a Lie-algebra valued one form $\displaystyle A = \frac{1}{2} A_{i}^{ab} dx^{i} \Gamma_{ab}$.
 The corresponding field strength $F$ is defined as usual, $F= dA +e A \wedge A$. Here $e$ is a gauge coupling constant.
Having finished with the description of our notation, we move to defining
the pseudo-energy, $E$, of the system,
\begin{align}
E&=   \frac{1}{16}\int {\rm Tr} \left\{ - F \wedge * F +  \alpha^2 (F\wedge F) \wedge
 *(F \wedge F)   \right\}~~. \label{pseudo-energy}
\end{align}
Here the word ``pseudo" means that this system is not an ordinary dynamical system because the space is a Riemannian space which has a positive definite metric $g_{ij}$  {without a time direction.} 
 {If we consider the space-time $\BR \times S^6$ 
by adding a time coordinate $\BR$, 
then $E$ can be regarded as a usual energy for static configurations. }
 {
In the case of the theory defined on a ten-dimensional space-time, the Riemannian space $S^6$ can be considered as the compactified space in 
$M^4 \times S^6$ with $M_4$ a pseudo Riemannian manifold 
\cite{Kihara:2007vz}.}  
The (mass) dimension of the gauge field is equal to $2$. The couplings $e$ and $\alpha$  have (mass) dimension $-1$ and $-3$, respectively.
 Thus this system is, of course, non-renormalizable.
We  leave the discussion of the physical meaning of our solution. 
We show the pseudo-energy in components expression:
\begin{align}
F &= \frac{1}{4} F_{ij}^{[ab]} dx^{ij} \Gamma_{ab}~,& \tr F \wedge * F &= -2 F_{ij}^{[ab]} F^{ij,[ab]} dv
\end{align}
The square of the Hodge star acts identically on these even forms, $**F=F~,**(F \wedge F)=F \wedge F$. 

The pseudo energy is  {bounded} 
from below by a topological charge with higher
derivative coupling with coupling constant $\alpha$,
\begin{align}
E &= \frac{1}{16} \int {\rm Tr}  \left( i F \mp \alpha \gamma_7 * (F \wedge F)
 \right) \wedge * \left( i F \mp\alpha 
 \gamma_7 * (F \wedge F)   \right) \pm \frac{i}{8}\alpha \int {\rm Tr}\gamma_7 F\wedge F\wedge
 F~~\cr
&\geq \pm \frac{i}{8}\alpha \int {\rm Tr}\gamma_7 F\wedge F\wedge F=  \mp \frac{1}{2^3}\int \epsilon_{abcdef} F^{[ab]} \wedge F^{[cd]} \wedge F^{[ef]} 
 \equiv \CQ~~,
\end{align}
where the pseudo-energy is bound by the topological number $\CQ$. 

The charge $\CQ$ is obtained by integrating a total derivative term and it reduces to a surface integral over a five-dimensional sphere 
 {as the boundary of one patch. 
We consider a trivial configuration 
for the field strength in the other patch. 
The two configurations of the two patches 
are transformed to each other by a gauge transformation 
in the overlap region (isomorphic to $S^5$).}
 Thus the integration is proportional to the winding number of $\Pi_5[{\rm SO(6)}]$.
 {The minimal group leading to a non-trivial fifth homotopy group is SU(3).}
 In general, the fifth homotopy group is closely related to non-abelian anomaly in four-dimensional space-time. 
The spin connection with respect to the standard metric of six-dimensional sphere provides an exact solution of the Tchrakian's generalized 
self-duality equation.
If the field strength fulfills the generalized (anti-)self-duality equation 
\begin{align}
F=\pm i \alpha \gamma_7 * (F \wedge F), \label{eq:gsdeq}
\end{align}
the pseudo-energy becomes $\CQ$. 
 {
Note that 
the product $F \wedge F$ and the Bogomol'nyi equation written by it 
depend on the representation matrices of $F$; 
The pseudo-energy and its Bogomol'nyi equation written 
by the SO(6) matrices are different from ours. 
}

Now we study an exact nontrivial solution of the self-duality (Bogomol'nyi) equation. We consider a hedge-hog connection $A$ and the field strength,
\begin{align}
A &= \frac{1}{4eR_0^2} x^a e^b \Gamma_{ab}  ~~,& F &=  \frac{1}{4eR_0^2} e^a \wedge e^b \Gamma_{ab}~~.
\end{align}
We can easily check that the field strength $F$ fulfills the {self-duality}
equation with $\alpha=eR_0^2/3$,
\begin{align}
F \wedge F 
&= - \frac{3}{e R_0^2} i \gamma_7 * F~~,& F &= \frac{eR_0^2}{3} i \gamma_7 * ( F \wedge F ) ~~.
\end{align}
Then the pseudo-energy is 
\begin{align}
E 
&= \frac{ 4\pi^3R_0^2}{e^2}~~. \label{eq:pseudo-energy}
\end{align}
This is a solution to the generalized self-duality equation 
(\ref{eq:gsdeq}).
 {The third power of the field strength gives a term proportional to volume form up to a numerical coefficient.}  The winding number of this solution is equal to 1.
We also obtained a solution with charge ${\cal Q}=-1$. The solution is given as $A^{(-)}:= \Gamma_6 A \Gamma_6$.
 The energy is given by the winding number of spatial $S^5$.  
Thus the spin connection with respect to the standard metric of six-dimensional sphere  {gives} an exact solution  {to} the Tchrakian's generalized 
self-duality equation (\ref{eq:gsdeq}). Our solution is the first one in 
($4p+2$)-dimensional theories with $p$ an integer. 
 {
It remains as an open question} 
whether the minimal group 
 needed for these solutions is SO(6) or not, 
since  
 {
the minimal group leading to a non-trivial 
fifth homotopy group is SU(3). 
}
The realization of this solution in a gauge theory coupled 
with gravity is a natural extension \cite{Kihara:2007vz}. 
It is needed in order to make contact with our universe.  
In particular the fact that the pseudo energy (\ref{eq:pseudo-energy}) 
is related to the radius $R_0$ of $S^6$ 
suggests that compactification from ten dimensions 
to four dimensions on $S^6$ with our solution
is stable without unwanted massless modes (moduli) \cite{Cremmer:1976ir}.
In addition, we can consider the similar system on another geometry. 
We have checked the case of complex projective space $\BC{\rm P}^3$. 
To make the Bogomol'nyi completion in the case of $\BC{\rm P}^3$, we must add another higher derivative coupling term which is not a single trace term.
Investigating possible relations 
with supergravity with higher derivative correction terms  
(see, e.g., \cite{Hyakutake:2006aq})
or with non-Abelian Dirac-Born-Infeld solitons 
is also an interesting subject.  

\begin{acknowledgments}
The authors would like to thank to Tigran Tchrakian for many advices and informing related references.
HK is grateful to Yutaka Hosotani, Hiroshi Itoyama, Yukinori Yasui and Takeshi Oota for useful comments.
Discussions with 
 Yoshifumi Hyakutake, Toshio Nakatsu and Asato Tsuchiya have been helpful to him.  
This work is supported by the 21 COE program ``Constitution of wide-angle mathematical basis focused on knots" from Japan Ministry of Education. 

\end{acknowledgments}


\begin{thebibliography}{99}
\bibitem{Belavin:1975fg}
  A.~A.~Belavin, A.~M.~Polyakov,  {A.~S.~Schwartz} and Yu.~S.~Tyupkin,
  Phys.\ Lett.\  B {\bf 59}, 85 (1975).

\bibitem{Seiberg:1994rs}
  N.~Seiberg and E.~Witten,
  Nucl.\ Phys.\  B {\bf 426}, 19 (1994)
  [Erratum-ibid.\  B {\bf 430}, 485 (1994)]
  [arXiv:hep-th/9407087];
  Nucl.\ Phys.\  B {\bf 431}, 484 (1994)
  [arXiv:hep-th/9408099].

\bibitem{Nordstrom:1988fi}
  G.~Nordstr\"om,
  Phys.\ Z.\  {\bf 15}, 504 (1914)
  [arXiv:physics/0702221].


\bibitem{Kaluza:1921tu}
  T.~Kaluza,
  Sitzungsber.\ Preuss.\ Akad.\ Wiss.\ Berlin (Math.\ Phys.\ ) {\bf 1921}, 966 (1921);
  O.~Klein,
  Z.\ Phys.\  {\bf 37}, 895 (1926)
  [Surveys High Energ.\ Phys.\  {\bf 5}, 241 (1986)].





\bibitem{brane-world}
    P.~Horava and E.~Witten, 
     Nucl.\ Phys.\ {\bf B460}, 506 (1996)  [arXiv:hep-th/9510209]; 
%
    N.~Arkani-Hamed, S.~Dimopoulos and G.~R.~Dvali,
     Phys.\ Lett.\ {\bf B429},  263 (1998) [arXiv:hep-ph/9803315]; 
    I.~Antoniadis, N.~Arkani-Hamed, S.~Dimopoulos and G.~R.~Dvali,
     Phys.\ Lett.\ {\bf B436},  257 (1998) [arXiv:hep-ph/9804398];
L.~Randall and R.~Sundrum, 
             Phys. Rev. Lett. {\bf 83}, 3370 (1999)  [arXiv:hep-ph/9905221]; 
             Phys. Rev. Lett. {\bf 83}, 4690 (1999)  [arXiv:hep-th/9906064].


\bibitem{Cremmer:1976ir}
  E.~Cremmer and J.~Scherk,
  Nucl.\ Phys.\  B {\bf 108}, 409 (1976);
  Nucl.\ Phys.\  B {\bf 118}, 61 (1977).


\bibitem{Tchrakian:1978sf}
  D.~H.~Tchrakian,
  J.\ Math.\ Phys.\  {\bf 21}, 166 (1980).

\bibitem{Tchrakian:1984gq}
  D.~H.~Tchrakian,
  Phys.\ Lett.\  B {\bf 150}, 360 (1985).

\bibitem{O'Se:1987fx}
  D.~O'Se and D.~H.~Tchrakian,
  Lett.\ Math.\ Phys.\  {\bf 13}, 211 (1987).

\bibitem{Ma:1990ja}
  Z.~Ma and D.~H.~Tchrakian,
  J.\ Math.\ Phys.\  {\bf 31}, 1506 (1990).

\bibitem{Grossman:1984pi}
  B.~Grossman, T.~W.~Kephart and J.~D.~Stasheff,
  Commun.\ Math.\ Phys.\  {\bf 96}, 431 (1984)
  [Erratum-ibid.\  {\bf 100}, 311 (1985)].

\bibitem{Chen:1999ab}
  B.~Chen, H.~Itoyama and H.~Kihara,
  Nucl.\ Phys.\ B {\bf 577}, 23 (2000)
  [arXiv:hep-th/9909075].

\bibitem{Kihara:2004yz}
  H.~Kihara, Y.~Hosotani and M.~Nitta,
  Phys.\ Rev.\ D {\bf 71}, 041701 (2005)
  [arXiv:hep-th/0408068].

\bibitem{Radu:2005rf}
  E.~Radu and D.~H.~Tchrakian,
  Phys.\ Rev.\ D {\bf 71}, 125013 (2005)
  [arXiv:hep-th/0502025].

\bibitem{Bergshoeff:1986jm}
  E.~Bergshoeff, M.~Rakowski and E.~Sezgin,
  Phys.\ Lett.\  B {\bf 185}, 371 (1987);
  A.~A.~Tseytlin,
  Nucl.\ Phys.\  B {\bf 501}, 41 (1997)
  [arXiv:hep-th/9701125];
  N.~E.~Grandi, E.~F.~Moreno and F.~A.~Schaposnik,
  Phys.\ Rev.\  D {\bf 59}, 125014 (1999)
  [arXiv:hep-th/9901073].


\bibitem{Kihara:2007vz}
  H.~Kihara and M.~Nitta,
  Phys.\ Rev.\  D {\bf 76}, 085001 (2007)
  [arXiv:0704.0505 [hep-th]].


\bibitem{Hyakutake:2006aq}
  Y.~Hyakutake and S.~Ogushi,
  JHEP {\bf 0602}, 068 (2006)
  [arXiv:hep-th/0601092].













\end{thebibliography}
\end{document}